     [2004/06/22 1.2 (KOF); 2001/04/25 1.1 (PWD)]

\documentclass[a4paper,twocolumn]{Gaia2004} 
\usepackage{times}   
\usepackage{epsfig}   
\usepackage{natbib}   
\def\mnras{MNRAS}
\def\apj{ApJ}

\def\aj{AJ}
\def\nat{Nature}
\def\apjl{ApJL}

\title{Substructure and Tidal Debris in Local Galaxies: Models and Observations}
\author[1,2]{James E.~Taylor}
\affil[1]{University of Oxford, Keble Road, Oxford OX1 3RH, United Kingdom}
\affil[2]{PPARC Fellow}

\bibpunct{(}{)}{;}{a}{}{,} 

\begin{document}


\maketitle

\begin{abstract}
One of the generic predictions of modern cosmological models is
that large galaxies should have experienced many mergers with 
smaller galaxies at some point in their past. Debris from such 
encounters will leave spatially distinct substructure in the 
stellar haloes of nearby galaxies, detectable for a few orbital 
periods after the final merger. In the case of the Milky Way, 
kinematic data from surveys such as RAVE and satellite missions such as 
GAIA will allow us to probe much more of the merger history, 
and to connect the properties of the stellar halo with those of 
local dwarf galaxies. To estimate what these programmes may discover, 
we review current observations of minor mergers in nearby galaxies, 
and compare these with predictions from a semi-analytic model of 
galaxy formation. 
\end{abstract}

\section{Introduction: The Role of Mergers In Current Cosmological Models}

In cosmological models consistent with current observations of the 
cosmic microwave background (CMB), large-scale structure, and estimates 
of the mean baryon density, roughly 85\% the matter in 
the universe is non-baryonic cold dark matter (CDM). Small
fluctuations in the initial distribution of CDM are known to lead, 
via gravitational instability, to the formation of dense, 
gravitationally bound structures at later times. These dark matter 
`haloes' should reach masses and densities sufficient to allow gas 
to cool and condense within them, thereby permitting efficient and 
sustained star formation, at redshifts or $\sim$20 or higher. Thus 
they represent a natural structural framework within which visible 
galaxies can form, from the earliest epoch onwards.

A generic prediction of cosmological models based on CDM is that the
dark matter haloes around galaxies continue to grow through accretion 
and mergers with smaller haloes, right up to the present day.
The merger rate for galaxy haloes reaches a peak at a redshift of 
$z\sim$1--3, however, and more recent mergers between galaxy haloes
and their associated galaxies will typically be minor ones, adding 
little to the total mass of the larger component. The average merger 
in the local universe may consist of a small, low surface-brightness 
dwarf galaxy being stretched out and disrupted over the course of several 
orbits around its parent. 
The only direct signature of such an event would be minor fluctuations 
in the surface brightness of the stellar halo around the larger galaxy, 
lasting for a few orbital periods (Johnston et al.~2001). 
Thus, while low-redshift mergers provide an important empirical test of 
the CDM framework, they may be extremely difficult to detect in practice.

In the case of the Milky Way, stellar kinematics provide a
unique opportunity to study the merger history in much greater
detail, identifying debris from older or more minor mergers 
thanks to its clustering in phase-space
(Tremaine 1993; Johnston 1998; Helmi \& White 1999). Ground-based
radial velocity surveys such as RAVE\footnote{http://www.aip.de/RAVE} 
and future astrometric
missions such as GAIA\footnote{http://astro.estec.esa.nl/GAIA} 
will explore a completely new range of
parameter space in the merger history of this one particular galaxy, 
and should thereby provide an important link between local stellar 
populations and the properties of small galaxies forming at
the highest redshifts. To explore the potential of these programmes, 
we will review current observations of debris from minor mergers in 
external galaxies, and attempt to extend these results to older or 
smaller satellites using a semi-analytic model of galaxy formation.

\section{Observed Streams In Nearby Galaxies}

Given their theoretical significance, there has been much 
recent interest in detecting ongoing minor mergers in local galaxies. 
The most spectacular discoveries of this kind were those made in
our own galaxy and in M31. In 1994, Ibata et al.~announced the
discovery of the Sagittarius Dwarf, a small system in the process of
being disrupted by the Milky Way, whose tidal debris is now spread
out over a large fraction of the sky. More recently, and equally
spectacular tidal stream was discovered around M31 (Ibata et al.\
2001; McConnachie et al.~2003), and there is growing evidence for 
another stream in the plane of the Milky Way
(e.g.~Newberg et al.~2002; Ibata et al.~2003). 
The LMC-SMC system can also be considered an ongoing
minor merger, since it has already experienced strong tidal effects
due to its proximity to the Milky Way (e.g.~Harris \& Zaritsky 2004), 
and will eventually merge with the Galaxy as dynamical friction 
acts on its orbit.

Much of the tidal debris detected locally has a very low surface brightness, 
and would be difficult or impossible to detect in distant
systems. Nonetheless, there has been some progress in detecting debris 
around galaxies outside the Local Group. Early work by Malin
\& Hadly (1997) found faint features in the stellar haloes of
several galaxies, including M83, M104 and NGC2855, some or all
of which may be tidal debris from minor mergers. A clear example 
of a tidal stream was discovered around the edge-on spiral NGC5907 
by Shang et al.~(1998), and the nearby active galaxy Centaurus~A has a
tidal feature of young blue stars, possibly associated with the recent
merger that triggered its central activity (Peng et al.~2002). These
results were reviewed and discussed recently by Pohlen et al.~(2004),
who also report newly detected streams in four other galaxies. Their
systematic search around 80 edge-on disk galaxies yielded only one of
these examples, however, suggesting the incidence of obvious streams
is low. More recently, Forbes et al.~(2004) have reported the
serendipitous discovery of a dwarf in the process of disruption, with
well-defined tidal streams, in the background of a Hubble Advanced
Camera image of the Tadpole system. These discoveries confirm that 
minor mergers are still taking place, but suggest that it may be hard to 
obtain a large sample to compare with theoretical models. Finally, we note 
that stellar streams are not the only tracer of minor mergers in external 
galaxies; others include distinct populations of globular clusters, gas 
or dust lanes in early-type galaxies, and kinematic or structural anomalies. 
The theoretical interpretation of these features is less straightforward, 
however, as their properties may depend more on the long-term response of 
the main system (e.g.~via AGN or starbursts) than on the nature of the initial
perturber.

\section{The Unresolved Problem of Dwarf Galaxy Formation}

Given this growing body of information on minor mergers and tidal streams, 
it is worth estimating how often these
events are expected to occur in current galaxy formation models. 
In recent years several groups have begun to study the formation of 
the stellar halo in its full cosmological context (e.g.~Bullock,
Kravtsov, \& Weinberg 2001; Bekki \& Chiba 2001; Brook et al.~2004; 
Bullock \& Johnston 2004). The problem is not
as straightforward as it seems, however. The properties of the stellar
halo will depend on when and how the satellites that merged into it
first formed. While simulations of structure formation have now
converged on the properties of dark matter structures on the relevant 
scales, it is still not clear how these structures are populated with 
visible stars.

It has been known since the early days of CDM that galaxy formation 
must be systematically less efficient in smaller
haloes, since the luminosity function of field galaxies flattens 
below some magnitude, whereas the halo mass function should be close 
to a power law over the corresponding range (e.g.~Kauffmann et al.~1993). More
recently, high-resolution  simulations by Klypin et al.~(1999) and
Moore et al.~(1999) have demonstrated that this discrepancy is even
more dramatic for satellites within systems like the Local
Group. Using internal velocities to relate dark matter subhaloes 
to visible systems, they showed that down to the scale of 
the smallest systems, there should be almost 100 times more subhaloes 
around the Milky Way than there are detected satellites. Subsequently, 
there has been much discussion of whether dwarf
galaxies populate a small subset of all subhaloes over a wide range of
mass/internal circular velocity, as proposed by Moore et al.~(1999),
or whether they correspond to kinematically cold stellar cores within 
the most massive subhaloes, as suggested by Stoehr et al.~(2002) and 
Hayashi et al.~(2003).

\begin{figure}[h]
 \begin{center} \leavevmode
\centerline{\epsfig{file=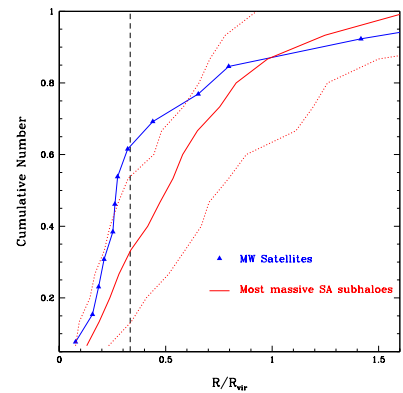, width=0.3 \textwidth}}
  \end{center} \caption{The cumulative radial distribution of visible
 satellites around the Milky Way (upper line/triangles), compared with the
 predicted distribution of the most massive subhaloes (lower
 solid line, with dotted 99\% contours.} \label{fig_1}
\end{figure}

The second of these solutions is appealing, as it could indicate that
there is a single characteristic halo mass below which star formation
ceases. This solution is strongly disfavoured, however, if not ruled out, 
by the strong spatial clustering of the satellites of the Milky
Way. The best estimates of the mass of the Milky Way put its virial
radius somewhere around 300 kpc. If this is the case, then two-thirds
of its satellites lie within the central third of the halo. The
chance of this happening if satellites populated only the most
massive haloes is less than 1\% (Taylor et al.~2003). 
The disagreement in the cumulative
radial distributions is shown in Figure~\ref{fig_1}, where we compare the
average results for the dozen most massive satellites from each of a
large set of semi-analytic models of halo substructure, generated
using the method outlined in Taylor \& Babul (2004a, 2004b, 2004c)
(solid line + dotted 99\% contours) with the distribution for the 
satellites of the Milky Way (upper line \& triangles). Thus at least 
some of the satellites must reside in smaller haloes.

\begin{figure}[h] \begin{center} \leavevmode
\centerline{\epsfig{file=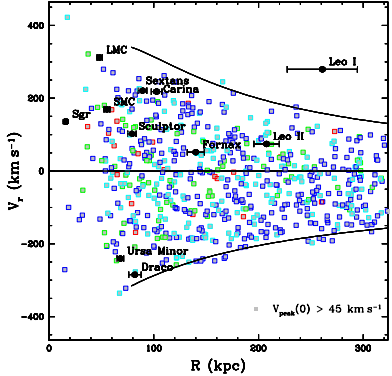, width=0.3 \textwidth}}
  \end{center} \caption{Radial velocity versus position for the
 satellites of the Milky Way (points with error bars) and the most
 massive subhaloes from a series of semi-analytic haloes (smaller symbols).}
 \label{fig_2}
\end{figure}

There may be additional clues to the origin of the 
local dwarfs in their kinematics, which appear very different from
those of randomly selected subhaloes. In Figure~\ref{fig_2}, for 
instance, we show how the positions and radial velocities of known 
satellites (points with error bars) compared with the distribution 
of massive haloes taken from a large set of semi-analytic haloes. 
There is a clear mis-match between the two distributions in 
projection along either axis. 

One possible explanation, both for the spatial clustering and for the observed 
velocity distribution, is that most of the dwarfs are old, and that 
the lower cutoff to dwarf formation increases with time. In particular, 
Kravtsov et al.~(2004) have produced a detailed model 
of dwarf formation where this occurs naturally, through a combination of 
internal and external feedback. In the next section we will consider 
a simplified version of this model and discuss its implications for 
stream formation.

\section{A Semi-analytic Model of Stream Formation} 
 
Given the complexity of modelling dwarf galaxy formation
self-consistently, it is useful to construct a simplified
semi-analytic model of the process, to get a preliminary estimate of
how often minor mergers with dwarf satellites would have occurred in
a system like the Milky Way. Our model is based on the semi-analytic
model of halo formation presented in (Taylor \& Babul 2004a,b,c),
which predicts the numbers, orbits, internal properties and dynamical 
evolution of dark matter subhaloes within a galaxy halo. To make
predictions about visible satellites, we will suppose that stars form
preferentially in subhaloes with deeper potential wells, but also
that the lower threshold to this process increases with time, such
that the some of the largest subhaloes around the Milky Way have never
formed stars. Specifically, we choose a time-varying threshold in peak
circular velocity of the form $V_{\rm p} > V_{\rm
p,0}\,(1+z)^{-\alpha}$. Subhaloes that exceed this threshold at any
time before they merge with the main halo are assumed to host visible
dwarf galaxies. The two parameters in this model can be adjusted to give 
the right total number and spatial distribution of surviving satellites at
the present-day.

Having identified which systems form stars, we can then estimate when
stellar material will be stripped from them and added to the stellar
halo. We assume the stars within a given subhalo are restricted to its
central region, in keeping with the observed sizes of dwarf galaxies,
and consequently that tidal stripping does not produce visible streams
until a system has lost most of its original dark matter. Considering
the stellar-to-total mass ratios of larger galaxies, we set a mass-loss
threshold at 83\%, and refer to systems that have lost less than this 
fraction of their original mass as `surviving satellites', and those 
that have lost more as `tidal streams'. Systems that lose 
more than 99\% of their mass may become completely unbound (Taylor \& 
Babul 2004a), so we refer to them as `disrupted' in what follows.

\begin{figure}[h] \begin{center} \leavevmode
\centerline{\epsfig{file=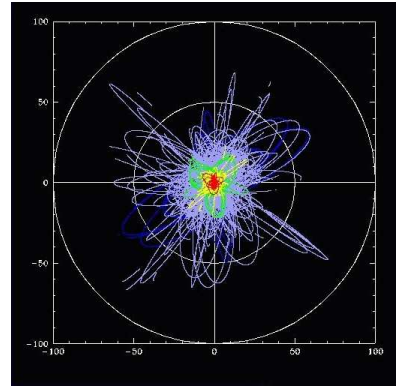, width=0.3 \textwidth}}
  \end{center} \caption{Tidal debris from merging satellites in a
 semi-analytic model halo, plotted at the point where it was first
 stripped. The material is colour-coded by age, the central material
 being systematically older. The figure is in the plane of 
 the disk and the scale is in kpc.} \label{fig_3}
\end{figure}

The distribution of stripped material from systems containing stars 
is shown in Figure \ref{fig_3}, for one particular semi-analytic halo. 
The debris is plotted at the point where it was first stripped off,
to highlight the underlying orbital structure.
Over time this material will be heated and phase-mixed, producing a
smoother final halo. The colours in the figure indicate the age at which 
the stripped system first merged with the main halo. 
There is a strong age gradient in the final system, with ancient mergers
contributing most of the material within the solar radius, while more
recent mergers add streams to the outer halo. 

\section{Results: Predicted Stream Properties}

\begin{figure}[h] \begin{center} \leavevmode
\centerline{\epsfig{file=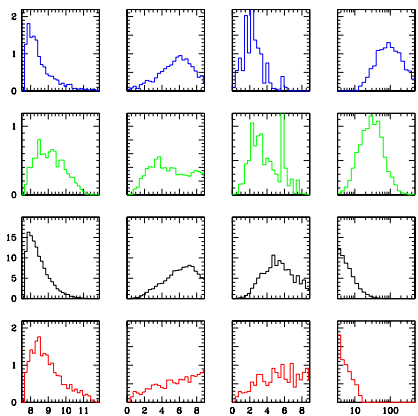, width=0.32 \textwidth}}
  \end{center} \caption{Distribution of original total mass,
 formation redshift, merger redshift, and final radial position
 (columns from left to right) of stellar systems that survive at the
 present-day, tidal stream progenitors, disrupted systems and central
 mergers (four rows, from top to bottom) in a large set of semi-analytic
 haloes.} \label{fig_4}
\end{figure}

Using the model outlined above, we have generated streams in a large
set of model haloes. We find that we can produce an average of ten
`surviving' satellites per halo, roughly the number known to-date for
the Milky Way, by setting $V_{\rm p,0} = 135$ km\,s$^{-1}$ and $\alpha
= 2/3$ in the expression given above. 
With this choice of parameters, our model predicts $\sim$8 `tidal
streams' (i.e.\ heavily stripped stellar systems) per halo, and
more than 100 disrupted dwarf galaxies, as well as 15 that have merged
directly into the centre of the main galaxy on very radial
orbits. Figure~\ref{fig_4} shows the distributions of various parameters,
including the original total mass, formation redshift, merger
redshift, and final radius (four columns, from left to right) for
these four classes of objects (four rows, from top to bottom).
Various interesting trends appear in these distributions; in particular,
the progenitors of tidal streams (second row) lie at somewhat smaller radii 
($\sim$ 30 kpc), are slightly older, and have larger original 
masses than the surviving satellites (top row). 

We can compare these predictions with the observations discussed
above. On the one hand, only a few streams have been detected in 
the Milky Way (Sagittarius, Canis, and possibly a few older streams 
associated with objects such as the massive globular cluster 
$\omega$ Cen; cf.~Bekki \& Freeman 2003).
On the other hand, given the difficulty of detecting older streams, 
substantial incompleteness is not implausible. Our results suggest 
that GAIA may expect to sample stars from a much larger population of 
disrupted systems -- on the order of 100. Most of these systems will 
have merged with the halo at redshifts of 2 or higher, however,
and will consist of old stars, strongly mixed in phase-space
by repeated scattering from their initial orbits. It will take much 
more detailed modelling of the disruption and mixing process 
(cf.~Johnston 1998; Helmi \& White 1999) to determine how many 
distinct structures should be detectable in practice.

\section*{Acknowledgements}

The author gratefully acknowledges helpful discussions with 
A.\ Babul, A.\ Ferguson, A.\ Kravtsov and J.\ Silk, as well as 
financial support from the U.K.\ Particle Physics and Astronomy 
Research Council (PPARC).

\end{document}